\documentclass{article}
\usepackage{amssymb}
\begin{document}
\title{Solutions of the Schr\"odinger equation given by solutions of the Hamilton--Jacobi equation}

\author{G.F.\ Torres del Castillo \\ Departamento de F\'isica Matem\'atica, Instituto de Ciencias \\
Universidad Aut\'onoma de Puebla, 72570 Puebla, Pue., M\'exico \\[2ex]
C.\ Sosa-S\'anchez \\ Facultad de Ciencias F\'isico Matem\'aticas \\ Universidad Aut\'onoma de Puebla, 72570 Puebla, Pue., M\'exico}

\maketitle

\begin{abstract}
We find the form of the potential depending on the coordinates and the time such that a solution, $S$, of the Hamilton--Jacobi equation yields an exact solution, $\exp ({\rm i} S/\hbar)$, of the corresponding Schr\"odinger equation.
\end{abstract}


\section{Introduction}
As is well known, the Hamilton--Jacobi (HJ) equation can be regarded as an approximation to the Schr\"odinger equation. More precisely, these two equations are related in the following way. If $\psi = \exp ({\rm i} S/ \hbar)$ is substituted into the Schr\"odinger equation for a particle of mass $m$ and electric charge $e$, subject to a velocity-independent potential $V$ and a magnetic field with vector potential ${\bf A}$,
\begin{equation}
- \frac{\hbar^{2}}{2m} \left( \nabla - \frac{{\rm i} e}{\hbar c} {\bf A} \right)^{2} \psi + V \psi = {\rm i} \hbar \frac{\partial \psi}{\partial t}, \label{sch}
\end{equation}
imposing the condition $\nabla \cdot {\bf A} = 0$, one obtains
\begin{equation}
\frac{1}{2m} \left( \nabla S - \frac{e}{c} {\bf A} \right)^{2} + V + \frac{\partial S}{\partial t} = \frac{{\rm i} \hbar}{2m} \nabla^{2} S, \label{equiv}
\end{equation}
which is equivalent to (\ref{sch}). Then, if one looks for a solution of this last equation of the form
\[
 S = S_{0} + \frac{\hbar}{\rm i} S_{1} + \left( \frac{\hbar}{\rm i} \right)^{2} S_{2} + \cdots,
 \]
one finds that $S_{0}$ satisfies the HJ equation
\begin{equation}
\frac{1}{2m} \left( \nabla S_{0} - \frac{e}{c} {\bf A} \right)^{2} + V + \frac{\partial S_{0}}{\partial t} = 0 \label{hj}
\end{equation}
(see, e.g., Ref.\ \cite{GY}, Sec.\ 2.8). However, Eq.\ (\ref{equiv}) shows that if $S$ satisfies the Laplace equation, $\nabla^{2} S = 0$, and the HJ equation (\ref{hj}), then $\psi = \exp ({\rm i} S/ \hbar)$ is an {\em exact}\/ solution of the Schr\"odinger equation (\ref{sch}).

The aim of this paper is to give some explicit examples of this relationship between solutions of the HJ equation and solutions of the corresponding Schr\"odinger equation. In Section 2 we show that from each solution, $S$, of the Laplace equation, and any given magnetic field, one can define a potential, $V$, in such a way that $S$ is a solution of the HJ equation and $\exp ({\rm i} S/ \hbar)$ is a solution of the corresponding Schr\"odinger equation. In Sections 3--5 we present some examples, restricting ourselves to problems without magnetic field. An example of this class was given in Ref.\ \cite{GF}, where the attention was focused on the HJ equation.

\section{Basic results}
According to Eq.\ (\ref{equiv}), assuming that $S$ is a solution of the Laplace equation
\begin{equation}
\nabla^{2} S = 0, \label{lap}
\end{equation}
$S$ satisfies the HJ equation (\ref{hj}) if and only if $\psi = \exp ({\rm i} S/ \hbar)$ is a solution of the Schr\"odinger equation (\ref{sch}). Instead of specifying the potentials $V$ and ${\bf A}$, and then try to find a simultaneous solution of the Laplace equation and the HJ equation, it is simpler to start from a solution, $S$, of the Laplace equation and {\em define}\/ the potential $V$ so that $S$ automatically satisfies the HJ equation (\ref{hj}). That is, to take
\begin{equation}
V = - \frac{\partial S}{\partial t} - \frac{1}{2m} \left( \nabla S - \frac{e}{c} {\bf A} \right)^{2} \label{potsv}
\end{equation}
or, if the magnetic field is absent, choosing ${\bf A} = {\bf 0}$,
\begin{equation}
V = - \frac{\partial S}{\partial t} - \frac{1}{2m} (\nabla S)^{2}. \label{pot}
\end{equation}
Of course, the resulting potential may be of little interest, but if one makes use of the general solution of the Laplace equation, one can find all the potentials for which a solution $S$ of the HJ equation corresponds to a solution $\exp ({\rm i} S/ \hbar)$ of the Schr\"odinger equation.

In the following sections we give some explicit examples, without magnetic field, starting from the highly illustrative one-dimensional case.

\section{Examples in one dimension}
In the one-dimensional case, the general solution of Eq.\ (\ref{lap}) is a linear function of the Cartesian coordinate $x$
\begin{equation}
S(x,t) = \alpha(t) \, x + \beta(t), \label{s1d}
\end{equation}
where $\alpha(t)$ and $\beta(t)$ are real-valued functions of $t$ only. Substituting (\ref{s1d}) into (\ref{pot}) we get
\begin{equation}
V = - \alpha'(t) \, x - \beta'(t) - \frac{1}{2m} \alpha^{2}(t), \label{pot1d}
\end{equation}
which corresponds to a possibly time-dependent uniform field of force. Apart from the trivial case of a free particle, two specific examples contained in (\ref{pot1d}) are a time-independent uniform field, and a uniform field with an intensity that changes linearly with time.

\subsection{Constant uniform field}
From Eq.\ (\ref{pot1d}) we see that in order to have a potential
\begin{equation}
V = - Fx, \label{pcu}
\end{equation}
where $F$ is a constant, corresponding to a constant uniform force of magnitude $F$, $\alpha'(t)$ must be equal to $F$ and $\beta'(t) = - \alpha^{2}(t)/2m$, hence
\[
\alpha(t) = Ft + P, \qquad \beta(t) = - \frac{(Ft + P)^{3}}{6mF},
\]
where $P$ is an arbitrary constant. Thus, substituting these expressions into (\ref{s1d}), we obtain
\begin{equation}
S(x,t) = (Ft + P) \, x - \frac{(Ft + P)^{3}}{6mF}. \label{cuf}
\end{equation}

The solution (\ref{cuf}) of the HJ equation corresponding to the potential (\ref{pcu}) is complete (by virtue of the presence of the arbitrary parameter $P$) and it is not separable (because of the term $Ftx$) and, for each value of $P$, the wavefunction
\begin{equation}
\psi(x,t) = \exp \frac{{\rm i}}{\hbar} \left[ (Ft + P) \, x - \frac{(Ft + P)^{3}}{6mF} \right] \label{scuf}
\end{equation}
is a solution of the Schr\"odinger equation with the potential (\ref{pcu}), which is not (multiplicatively) separable (and is not a stationary state). The wavefunctions (\ref{scuf}) are non-normalizable, just as the ``plane waves'' $\exp ({\rm i} P x/\hbar)$, and, as the latter, for each value of $t$, form a complete set, in terms of which any wavefunction can be expanded (by means of an integral on the parameter $P$). In fact, in the same manner as the plane waves are eigenfunctions of the momentum operator, the wavefunctions (\ref{scuf}) are eigenfunctions of the (time-dependent) operator $p - Ft$ (which is a conserved quantity), with eigenvalue $P$ \cite{TD}.

\subsection{Field with constant growth}
We now consider the time-dependent potential
\begin{equation}
V(x) = - kt x, \label{fug}
\end{equation}
where $k$ is a constant. This potential coincides with (\ref{pot1d}) if $\alpha'(t) = kt$ and $\beta'(t) = - \alpha^{2}(t)/2m$, hence
\[
\alpha(t) = \frac{kt^{2}}{2} + P, \qquad \beta(t) = - \frac{1}{2m} \left( \frac{k^{2} t^{5}}{20} + \frac{kPt^{3}}{3} + P^{2} t \right),
\]
where $P$ is an arbitrary constant. Thus,
\begin{equation}
S(x,t) = \left( \frac{kt^{2}}{2} + P \right) x  - \frac{1}{2m} \left( \frac{k^{2} t^{5}}{20} + \frac{kPt^{3}}{3} + P^{2} t \right) \label{cg}
\end{equation}
is a complete solution of the HJ equation with potential (\ref{fug}) and
\begin{equation}
\psi(x,t) = \exp \frac{{\rm i}}{\hbar} \left[ \left( \frac{kt^{2}}{2} + P \right) x  - \frac{1}{2m} \left( \frac{k^{2} t^{5}}{20} + \frac{kPt^{3}}{3} + P^{2} t \right) \right] \label{scg}
\end{equation}
is a solution of the corresponding Schr\"odinger equation.

The (non-normalizable) wavefunctions (\ref{scg}) form a complete set. In fact, (\ref{scg}) is eigenfunction of the conserved operator $p - {\textstyle \frac{1}{2}} k t^{2}$, with eigenvalue $P$.

\section{Examples in two dimensions}
The solutions of the Laplace equation in two dimensions are the form
\begin{equation}
S(x,y,t) = {\rm Re} [f(z,t)], \label{anf}
\end{equation}
where $z \equiv x + {\rm i} y$, and $f(z,t)$ is an analytic function of $z$ that may also depend on $t$ and on other parameters. Substituting (\ref{anf}) into Eq.\ (\ref{pot}) we obtain the expression for the potential
\begin{equation}
V = - \frac{1}{2} \frac{\partial (f + \overline{f})}{\partial t} - \frac{1}{2m} \left| \frac{\partial f}{\partial z} \right|^{2}. \label{pot2d}
\end{equation}

In the following subsections we give two explicit examples corresponding to specific choices of $f$.

\subsection{The repulsive isotropic harmonic oscillator}
The function
\[
f(z,t) = \frac{m \omega}{2} z^{2} + (P_{1} {\rm e}^{- \omega t} - {\rm i} P_{2}  {\rm e}^{\omega t}) \, z + \frac{1}{4 m \omega} (P_{1} {\rm e}^{- \omega t} - {\rm i} P_{2}  {\rm e}^{\omega t})^{2},
\]
where $m$, $\omega$, $P_{1}$, and $P_{2}$ are constants, is an analytic function of $z$ and, from Eq.\ (\ref{pot2d}), we obtain the potential
\begin{equation}
V = - \frac{m \omega^{2}}{2} (x^{2} + y^{2}), \label{potrho}
\end{equation}
which corresponds to a ``repulsive harmonic oscillator'' (see, e.g., Ref.\ \cite{BW}). Then, making use of Eq.\ (\ref{anf}), we have the complete solution of the HJ equation
\begin{equation}
S(x,y,t) = \frac{m \omega}{2} (x^{2} - y^{2}) + P_{1} \, {\rm e}^{- \omega t} x + P_{2} \, {\rm e}^{\omega t} y + \frac{1}{4 m \omega} (P_{1}{}^{2} \, {\rm e}^{- 2 \omega t} - P_{2}{}^{2}  \, {\rm e}^{2 \omega t}) \label{hjrho}
\end{equation}
and the solutions of the Schr\"odinger equation
\begin{equation}
\psi(x,y,t) = \exp \frac{{\rm i}}{\hbar} \left[ \frac{m \omega}{2} (x^{2} - y^{2}) + P_{1} \, {\rm e}^{- \omega t} x + P_{2} \, {\rm e}^{\omega t} y + \frac{1}{4 m \omega} (P_{1}{}^{2} \, {\rm e}^{- 2 \omega t} - P_{2}{}^{2}  \, {\rm e}^{2 \omega t}) \right]. \label{srho}
\end{equation}
Each wavefunction (\ref{srho}) is a common eigenfunction of the conserved operators ${\rm e}^{\omega t} (p_{x} - m \omega x)$ and ${\rm e}^{- \omega t} (p_{y} + m \omega y)$ with eigenvalues $P_{1}$ and $P_{2}$, respectively.

\subsection{The central potential $V = -k/r^{2}$}
The function
\[
f(z,t) = \sqrt{2mk} \, \ln z,
\]
where $k$ is a constant, is an analytic function of $z$ (except at $z = 0$) and leads to the central potential
\begin{equation}
V(x,y) = - \frac{k}{x^{2} + y^{2}}. \label{invs}
\end{equation}
Thus,
\[
S = \sqrt{\frac{mk}{2}} \ln (x^{2} + y^{2})
\]
is a solution of the HJ equation with the potential (\ref{invs}). By contrast with the previous examples, this solution does not contain arbitrary parameters and does not depend on the time.

\section{Examples in higher dimensions}
Apart from the potentials that can be defined directly with the aid of Eq.\ (\ref{pot}), making use of a solution of the Laplace equation, as in the foregoing sections, we can also combine the potentials already obtained in low dimensions to construct potentials in two or more dimensions. For instance, from Eqs.\ (\ref{cuf}) and (\ref{hjrho}) we can form the function
\[
S(x,y,z,t) = \frac{m \omega}{2} (x^{2} - y^{2}) + P_{1} \, {\rm e}^{- \omega t} x + P_{2} \, {\rm e}^{\omega t} y + \frac{1}{4 m \omega} (P_{1}{}^{2} \, {\rm e}^{- 2 \omega t} - P_{2}{}^{2}  \, {\rm e}^{2 \omega t}) + (Ft + P_{3}) \, z - \frac{(Ft + P_{3})^{3}}{6mF}
\]
which satisfies the Laplace equation in three dimensions and leads to the potential [cf.\ Eqs.\ (\ref{pcu}) and (\ref{potrho})]
\[
V(x,y,z) = - \frac{m \omega^{2}}{2} (x^{2} + y^{2}) - Fz
\]
and, therefore, $\exp ({\rm i} S/\hbar)$ is a solution of the Schr\"odinger equation with this last potential, which contains three quantum numbers $P_{1}, P_{2}$, and $P_{3}$.

\section{Concluding remarks}
It may be remarked that, for a given potential, the existence of a solution of the HJ equation that also satisfies the Laplace equation, does not imply that all solutions of this HJ equation will satisfy the Laplace equation. In the examples of Section 3, the principal functions (\ref{cuf}) and (\ref{cg}) (which contain an arbitrary parameter, $P$) are, up to an additive constant, the {\em only}\/ common solutions of the corresponding HJ equation and the Laplace equation.

One open question is what are the physical or geometrical properties of a potential such that the corresponding HJ equation admits solutions that also satisfy the Laplace equation. Another question is the meaning of the quantum states obtained in this manner.

The eikonal equation of geometrical optics can be regarded as an approximation to the scalar Helmholtz equation, in a similar manner as the HJ equation is an approximation to the Schr\"odinger equation and, therefore, one can expect that, in some cases, a solution of the eikonal equation would produce an exact solution of the Helmholtz equation. This connection is currently under investigation \cite{GS}.

\section*{Acknowledgements}
The authors acknowledge Dr.\ Gilberto Silva Ortigoza for useful comments. One of the authors (C.S.S.) wishes to thank the Consejo Nacional de Ciencia y Tecnolog\'ia for financial support.

\end{document}